\documentstyle[12pt]{article}
\renewcommand{\theequation}{\thesection.\arabic{equation}}
\newcommand{\beq}{\begin{equation}}
\newcommand{\eeq}{\end{equation}}
\newcommand{\bea}{\begin{eqnarray}}
\newcommand{\eea}{\end{eqnarray}}

\begin{document}
\setcounter{page}{0}
\topmargin 0pt
\oddsidemargin 5mm
\renewcommand{\thefootnote}{\fnsymbol{footnote}}
\newpage
\setcounter{page}{0}
\begin{titlepage}

\begin{flushright}
QMW-PH-97-21\\
{\bf hep-th/9707043}\\
 {\it July 1997}
\end{flushright}
\vspace{0.5cm}
\begin{center}
{\Large {\bf Random lattice strings vs. type IIB matrix models}} \\
\vspace{1.8cm}
\vspace{0.5cm}
{ Oleg A.
Soloviev\footnote{e-mail: O.A.Soloviev@QMW.AC.UK}} \\
\vspace{0.5cm}
{\em Physics Department, Queen Mary and Westfield College, \\
Mile End Road, London E1 4NS, United Kingdom}\\
\vspace{0.5cm}
\renewcommand{\thefootnote}{\arabic{footnote}}
\setcounter{footnote}{0}
\begin{abstract}
{I comment on a curious relation between Siegel's model of random lattice strings and type %%@
IIB matrix model. The comparison of the two theories suggests that there may exist extra %%@
terms in the latter which are overlooked in the weak string coupling limit.
}
\end{abstract}
\vspace{0.5cm}
%\centerline{July 1996}
 \end{center}
\end{titlepage}
\newpage
%******************************************************************
\renewcommand{\theequation}{\arabic{equation}}
\setcounter{equation}{0}

Various matrix models \cite{Banks},\cite{Ishibashi},\cite{Fayyazuddin} which are now %%@
extensively discussed in the context of non-perturbative string theory do not necessarily %%@
refer to discretization of the world sheet. Matrices in these theories emerge as images of %%@
non-commuting coordinates associated with either D-particles \cite{Banks} or D-instantons %%@
\cite{Ishibashi},\cite{Fayyazuddin}. In other words, the integration over random matrices in %%@
these models is not intending to replace the summation over world sheet topologies. At the %%@
same time the random matrix approach based on the idea of a discretization of the world sheet %%@
has proved remarkably successful in understanding the non-perturbative structure of string %%@
theories in dimensions less than two \cite{old}. Therefore, it might be interesting to look %%@
for some possible interplay between new and old matrix model approaches. One nontrivial %%@
example of such a relation has already been discussed in \cite{Kristjanson}.

In the present letter, I would like to focus on one remarkable fact about random lattice %%@
strings which may turn out to be relevant in current attempts to construct a non-perturbative %%@
string theory. It has been observed that planar Feynman diagrams of matrix models have a %%@
duality symmetry under replacing vertices with loops and vice versa  %%@
\cite{David},\cite{Boulatov}. In \cite{Siegel1} this symmetry of random lattices has been %%@
connected to the T-duality of long- and short-distance behavior in string theory %%@
\cite{Kikkawa}. Namely, the duality invariance of planar Feynman diagrams gives rise to the %%@
T-duality invariance of string perturbation theory. 

Siegel was the first to realize that the T-duality of the continuum string in its random %%@
lattice representation uniquely determines the random matrix model potential \cite{Siegel2}.
His approach has further been generalized in \cite{Kuroki} to include self-dual matter %%@
systems.

Siegel's T-self-dual matrix model is given as follows \cite{Siegel2}
\begin{equation}
S_\Phi=\mbox{tr}\left[{1\over2}\Phi^2~+~N\;\ln(1-g\Phi)\right],\label{D=0}\end{equation}
where $\Phi$ is a hermitian $N\times N$ matrix and $g$ is a constant. This model can be %%@
thought of as describing a $D=0$ string.

Let us make the following change of variables
\begin{equation}
1-g\Phi=gY.\label{change}\end{equation}
In terms of $Y$ eq.(\ref{D=0}) takes the following form
\begin{equation}
S_Y=\mbox{tr}\left[-{Y\over g}~+~N\;\ln Y~+~{Y^2\over2}~+~{1\over2g^2}~+~N\;\ln %%@
g\right].\label{Y}\end{equation}

In the limit $g\to0$, $N\to\infty$ eq.(\ref{Y}) goes to
\begin{equation}
S_Y(g\to0,N\to\infty)=\mbox{tr}\left[-{Y\over g}~+~N\;\ln %%@
Y~+~constant\right].\label{limit}\end{equation}
The constant term is irrelevant for the duality property. The limit $g\to0,~N\to\infty$ is %%@
taken so that $g\cdot N$ is kept fixed.

The curious fact is that, up to a normalization of the constants $g$ and $N$, %%@
eq.(\ref{limit}) coincides with the saddle point of the type IIB matrix model %%@
\cite{Kristjanson}. According to ref.\cite{Fayyazuddin}, the constant $g$ in front of the %%@
linear in $Y$ term is proportional to the string coupling constant, $g\sim g_s$. Therefore, %%@
the matrix action given by eq.(\ref{limit}) can be considered as a weak string coupling limit %%@
of the Siegel matrix model. In particular, the latter has to contain type IIB D-branes which %%@
are observed among solutions of the IIB matrix model \cite{Ishibashi},\cite{Fayyazuddin}.

However, if we take the Siegel model as a definition of the exact type IIB matrix theory at %%@
the saddle point, then the $Y^2$-term in eq.(\ref{Y}) becomes important at finite values of %%@
the string coupling constant. It is natural to assume that the duality invariance of the %%@
Siegel random matrix model is an underlying symmetry of string perturbation theory and as %%@
such has to be preserved in any non-perturbative formulation. Bearing this principle in mind, %%@
one can write down a modified type IIB matrix model. Namely,
\begin{equation}
S_{IIB}=\alpha\;\mbox{tr}\left\{-{1\over4}Y^{-1}[A_\mu,A_\nu]^2~-%%@
~{1\over2}(\bar\psi\Gamma^\mu[A_\mu,\psi])\right\}~+~\mbox{tr}\left[\beta Y+\gamma\:\ln %%@
Y+\sigma Y^2+\delta\right].\end{equation}
Here $A_\mu$ and $\psi_\alpha$ are $N\times N$ hermitian bosonic and fermionic matrices %%@
respectively. This theory possesses the N=2 supersymmetry in the limit $N\to\infty$ %%@
\cite{Fayyazuddin}. The constant parameters $\alpha,~\beta,~\gamma,~\sigma,~\delta$ get, in %%@
general, renormalized. The hope is that they run to the values predicted by the T-duality. %%@
This will be studied elsewhere.

To conclude, the Siegel random lattice string and the type IIB matrix model have quite %%@
different setups. However, both theories aim at one and the same target - a non-perturbative %%@
description of string theory (whatever it might be). Therefore, it is not very surprising if %%@
the two approaches eventually will merge. 

I thank PPARC for financial support.

\end{document}